\documentclass[a4paper,11pt,showpacs,showkeys]{revtex4}
\usepackage[dvips]{graphicx}
\usepackage{amsfonts,amssymb,amsmath}
\newcommand {\be}{\begin{eqnarray}}
\newcommand {\ee}{\end{eqnarray}}

\begin{document}
\medskip
\title{Particle multiplicities and particle ratios in excluded volume model}

\author{M. Mishra\footnote{Email: madhukar.12@gmail.com} and C. P. Singh\footnote{Email: cpsingh\_bhu@yahoo.co.in}} 
\vskip 2in
\affiliation{Department of Physics, Banaras Hindu University,
Varanasi-221 005, India}

\begin{abstract}
     One of the most surprising results is to find that a consistent description of all the experimental results on particle multiplicities and particle ratios obtained from the lowest AGS to the highest RHIC energies is possible within the framework of a thermal statistical model. We propose here a thermodynamically consistent excluded-volume model involving an interacting multi-component hadron gas. We find that the energy dependence of the total multiplicities of strange and non-strange hadrons obtained in this model agrees closely with the experimental results. It indicates that the freeze out volume of the fireball is uniformly the same for all the particles. We have also compared the variation of the particle ratios such as $\langle K^{+}\rangle/\langle\pi^{+}\rangle,\;\langle K^{-}\rangle/\langle\pi^{-}\rangle,\; K^{-}/K^{+},\;\bar{p}/p,\; \bar{\Lambda}/\Lambda,\;\bar{\Xi }/\Xi,\;\bar{\Omega}/\Omega,\;\langle \Lambda\rangle/\langle\pi\rangle,\;\langle\Xi^{-}\rangle/\langle\pi\rangle,\;\langle\Omega+\bar{\Omega}\rangle/\langle\pi\rangle$ and $\langle\Phi\rangle/\langle\pi\rangle$ with respect to the center-of-mass energy as predicted by our model with the recent experimental data. 
\end{abstract}   

\pacs{:25.75.–q,; 21.65.+f; 24.10.Pa}
\keywords{:Heavy-ion collisions; multiplicity; strangeness, particle ratios; thermal model.}
\maketitle

\newpage
  
\section{Introduction}
 In ultra-relativistic heavy-ion collision, we expect that a matter with high energy density will be produced and finally a colour-deconfined matter is likely to be produced. Such a matter, better known as quark-gluon plasma (QGP), is only a transient state of the system because evolution back to ordinary hadronic matter begins immediately~\cite{satz,cps,cps1}. Consequently we face a difficult question how to diagnose and probe the plasma state. One of the most potential probe for detecting QGP formation is the observation of the abundance of the strange particles. This is based on the proposal that in the baryon rich plasma, the strange quark-antiquark pairs $s\bar s$ would be more abundant than u, d quarks because of the existence of Pauli blocking in light quark creation. The threshold for $s\bar s$ pair production ($\approx 300$ MeV) in QGP is much lower than that for $K\bar K$ production ($\approx 980$ MeV) in HG. Based on these arguments, Rafelski and M\"{u}ller~\cite{jraph,raph1,raph2,raphm,kom,kom1,egg} have suggested that a large strange quark present in QGP would automatically yield an enhanced particle production in the HG resulting after hadronization of QGP. However, one important assumption in the theory is that the HG formed after QGP phase does not find sufficient time to achieve chemical equilibrium, otherwise the equilibrated HG will not retain the memory regarding the primordial phase that might have existed earlier in the evolution process. However, in the analysis of strangeness enhancement, one should have the precise knowledge not only of the equation of state (EOS) of QGP alone but of HG also which gives the background contribution in this case.

  In ultra-relativistic nucleus-nucleus collisions, a very high density of matter exists over an extended region, and it is often called a 'fireball'. The physical variables characterizing a fireball are the energy density $\epsilon$, the baryon density $n_B$ and the volume. Thus one of the central problems in the study of high energy collisions lies in deducing the state of the system by determining temperature and baryon chemical potential $\mu_B$ (which in turn determine $\epsilon$ and $n_B$) existing in the fireball from the observed final particle multiplicities etc. The freeze-out stage of the system when the particles fly towards the detectors without further interactions is directly connected to the observed multiplicity distributions of various particles. Here we should emphasize that the fireball goes through two types of freeze-out stages. First it suffers a chemical freeze-out when inelastic collisions in the fireball do not occur anymore. Later when elastic collisions also stop in the fireball, the stage is called as thermal freeze-out. Abundances of the particles and their ratios provide important information regarding the chemical equilibrium occurring in the fireball before the thermal equilibrium. Moreover, chemical equilibrium in the hot, dense HG removes any memory existing in the fireball regarding a primordial phase transition. So the thermodynamical properties of the fireball do not throw any light on the existence of QGP before hadronization. We first connect the thermodynamical properties of the fireball to an appropriate EOS for the hot and dense hadron gas and finally we deduce the chemical freeze-out conditions of the thermal HG fireball formed in the ultra-relativistic heavy-ion collisions. In the last few years, one has observed a very surprising result. It has been shown that a consistent and appropriate description of all the experimental results on the particle multiplicities and particle ratios from the lowest SIS to the highest RHIC energies is available within the framework of a thermal statistical model as applied to a hot and dense HG phase. Recently we have given a model based on the geometrical excluded volume correction which describes suitably the thermodynamical quantities of a hot and dense HG~\cite{cp}. We have further used this prescription to determine the freeze-out volume of the fireball as well as pions and nucleon density~\cite{cp1}. We find that the densities are reproduced well by their HBT experimental data.  

   In this paper we attempt to explain the experimental data on strange particle multiplicities, particle ratios in the HG scenario using our excluded volume model discussed in~\cite{cp,cp1}. Since many attempts~\cite{and3} have been made by various authors to explain particle multiplicities using different HG models~\cite{cleyman,cleym,cley,braun,brau}, it will be worthwhile to compare chemical freeze-out parameters (i.e., $T,\,\mu_B$) of the fireball as determined in these models. Finally we show our curves for strange particle multiplicities and particle ratios and compare them with the experimental data. We also show the predictions of the other important models existing in the literature~\cite{cleym,jcley06,drisch}. It is indeed encouraging to notice that simple thermal models can explain well the experimental data on particle multiplicities and the strangeness enhancement. But the purpose here is to demonstrate that the whole exercise crucially depends on the EOS used to determine the thermodynamic state of the fireball of the hot and dense HG.                 
\section{Method of Calculation}
   We have formulated a new thermodynamically consistent excluded-volume model~\cite{cp} for the description of hot and dense hadron gas (HG). Here we are describing in brief for the sake of completeness. This model assumes that the baryon of $i^{th}$ species have an eigen volume $V_i$. If $R=\sum_i n_i^{ex}\,V_i$ be the fraction of occupied volume, then the number density $n_i^{ex}$ of $i^{th}$ baryon can be written as:
\begin{equation}
n_i^{ex}=(1-R)\,I_i\,\lambda_i-I_i\,\lambda_i^2\,\frac{\partial R}{\partial \lambda_i},
\end{equation} 
where $\lambda_i$ is the fugacity of $i^{th}$ baryons and $I_i$ is the following expression containing modified Bessel function of second kind
\begin{equation}
I_i=\frac{g_i}{2\pi^2}\left(\frac{m_i}{T}\right)^2\,T^3\,K_2(m_i/T)
\end{equation}
with $g_i$ is the spin-isospin degeneracy factor. Eq. (1) can be rewritten in the form
\begin{equation}
R=(1-R)\sum_i\,n_i^0\,V_i-\sum_i\,n_i^0\,V_i\,\lambda_i\,\frac{\partial R}{\partial \lambda_i},
\end{equation} 
with $n_i^0=I_i\,\lambda_i$. Taking $R^0=\sum_i\,X_i$, where $X_i=I_i\,\lambda_i\,V_i$   and putting $\partial R/\partial \lambda_i=0$, we get
\begin{equation}
R=\hat R=\frac{R^0}{1+R^0}.
\end{equation}
Thus we can write Eq. (3) in the form
\begin{equation}
R= \hat R+\Omega\, R,
\end{equation}
with $\Omega$ is given by
\begin{equation}
\Omega=-\frac{1}{1+R^0}\,\sum_i\,I_i\,\lambda_i^2\,V_i\,\frac{\partial}{\partial \lambda_i}.
\end{equation}
Now by following Neumann iteration method Eq. (3) can be cast in the form
\begin{equation}
R=\hat R+\Omega \hat R+\Omega^2\hat R+\Omega^3\hat R+\cdots
\end{equation}
Retaining terms upto second order only, the expression for $R$ can be written as
\begin{eqnarray}
R=\frac{\sum_i\,X_i}{1+\sum\,X_i}-\frac{\sum_i\,X_i^2}{(1+\sum_i\,X_i)^3}+
2\frac{\sum_i\,X_i^3}{(1+\sum_i\,X_i)^4}\\\nonumber
-3\frac{\sum_i\,X_i\,\lambda_i\sum_i\,
X_i^2\,I_i\,V_i}{(1+\sum_i\,X_i)^5}.
\end{eqnarray}
                                        
Finally by calculating the values of $R$ and $\partial R/\partial \lambda_i$, one can calculate the value of particle number density by using Eq. (1). 

We have considered here a hot and dense hadron gas with baryonic and mesonic resonances having masses up to 2 GeV/c$^2$. In order to conserve strangeness quantum number, we have used the criteria of net strangeness number density equal to zero. In all the above calculations we have also considered that all the mesons behave as point-like particles. Furthermore, we have taken equal volume $V=4\pi\,r^3/3$ for all baryons with a hard-core radius $r=0.8$ fm. 
\begin{table}[h!]
\caption{List of particle ratios used to fit different models~\cite{and3}.}
\begin{center}
\begin{tabular}{|c|c|c|c|}
\hline\hline
$\sqrt{s}$ (GeV) & Experiments & Particle ratios\\
\hline
2.70                   & AGS & $K^{+}/\pi^{+}$\\
                       &     & $p/\pi^{-}$\\
                       &     & $\Lambda/\pi^{-}$\\ 
3.32, 3.84, 4.30       & AGS & $K^{-}/K^{+}$\\
                       &     & $K^{+}/\pi^{+}$\\
                       &     & $K^{-}/\pi^{-}$\\
                       &     & $p/\pi^{-}$\\
                       &     & $\Lambda/\pi^{-}$\\ 
4.85                   & AGS & $K^{-}/K^{+}$\\
                       &     & $K^{+}/\pi^{+}$\\
                       &     & $K^{-}/\pi^{-}$\\
                       &     & $p/\pi^{-}$\\
                       &     & $\Lambda/\pi^{-}$\\ 
                       &     & $\bar p/p$\\
                       &     & $\bar \Lambda/\Lambda$\\ 
8.76                   & SPS & $K^{-}/K^{+}$\\
                       &     & $\bar\Lambda/\Lambda$\\
                       &     & $K^{+}/\pi^{+}$\\
                       &     & $K^{-}/\pi^{-}$\\
                       &     & $\bar\Lambda/\pi^{-}$\\                       &     & $\Xi/\pi^{-}$\\
                       &     & $\Xi/\Lambda$\\
                       &     & $\Omega/\Xi$\\
12.3                   & SPS & $K^{-}/K^{+}$\\
                       &     & $\bar \Lambda/\Lambda$\\                       &     & $K^{+}/\pi^{+}$\\
                       &     & $K^{-}/\pi^{-}$\\
                       &     & $\bar\Lambda/\pi^{-}$\\
\hline\hline
\end{tabular}
\end{center}
\end{table}

\begin{table}[h!]
\begin{center}
\begin{tabular}{|c|c|c|c|}
\hline\hline
$\sqrt{s}$ (GeV) & Experiments & Particle ratios\\
\hline
17.3                   & SPS & $K^{-}/K^{+}$\\
                       &     & $\bar\Lambda/\Lambda$\\
                       &     & $\bar\Xi/\Xi$\\
                       &     & $\bar\Omega/\Omega$\\
                       &     & $K^{+}/\pi^{+}$\\
                       &     & $K^{-}/\pi^{-}$\\
                       &     & $\bar\Lambda/\pi^{-}$\\ 
                       &     & $\Xi/\pi^{-}$\\
                       &     & $\Omega/\pi^{-}$\\
130, 200               & RHIC& $K^{-}/K^{+}$\\
                       &     & $\bar\Lambda/\Lambda$\\
                       &     & $\bar\Xi/\Xi$\\
                       &     & $\bar\Omega/\Omega$\\ 
                       &     & $K^{+}/\pi^{+}$\\
                       &     & $K^{-}/\pi^{-}$\\
                       &     & $\bar p/p$\\
                       &     & $p^{-}/\pi^{-}$\\
                       &     & $\bar\Lambda/\pi^{-}$\\
                       &     & $\Xi/\pi^{-}$\\
                       &     & $\Omega/\pi^{-}$\\ 
\hline\hline 
\end{tabular}
\end{center}
\end{table}  
    We consider all hadronic resonances having well defined masses i.e., their decay widths are small. All hadronic resonances decay rapidly in strong decays after freeze-out and thus contribute to the stable particle abundances. Some heavy resonances may decay in cascades. This has been implemented in the calculation by considering all decays proceeding  sequentially from the heaviest to lightest particles. As a consequence of this, the light particles get contributions from the heaviest particles. Thus it has the form :
\begin{eqnarray*}
n_1&=&b_{2\rightarrow1}\cdots b_N\,b_{N-1}\,n_N,                             
\end{eqnarray*}                                                       
where $b_{k\rightarrow k-1}$ combines the branching ratios for the $k\rightarrow k-1$ decay~\cite{eidel} with the appropriate Clebsch-Gordan Coefficients. The latter accounts for the isospin symmetry in strong decays and allows us to treat separately the different charged states of isospin multiplets of particles. In order to calculate contributions from some heavy resonance decays we are forced to take some approximations. For example, in most of the cases there are several decay channels. In our approach we discard all decays with branching ratios less than 2\%. In addition if the decay channels are classified as dominant, large seen or possibly seen, we take into account the dominant channel only. If two or more than two channels are described as equally important, we take all of them with the same weight. For example, $f_0(980)$ decays into $\pi$ (according to~\cite{eidel} this is the dominant channel) and $K\bar K$ (according to~\cite{muller} this is the seen channel). In our approach, as a rule stated above, we include only the process $f_0(980)\rightarrow \pi$. Similarly $a_0(1450)$ has three decay channels: $\eta\,\pi$ (seen), $\pi\,\eta^{'}$ (958) (seen) and $K\bar K$ (again seen). In this case we include all the three decay channels with equal weight $0.33$ (branching ratios). Of course, this procedure is not unique and may vary from author to author.  
   Another problem that we face in this calculation is that in some decay channels the branching ratios are not given exactly but a range of values are given and the sum of the branching ratios may differ significantly from $1.0$. In this case we take the mean values of the branching ratios. Since we require that their sum should be properly normalized, we are forced to rescale all the mean values in such a way that their sum is indeed $1.0$. Since the experimental data on the resonances involve a lot of uncertainties in decay width as well as in branching ratios, we adopt the above procedure in calculating the contributions of resonances towards a given particle multiplicity. Unfortunately, a better and established alternative for such calculation does not exist in the literature.  
 \begin{table}[h!]
\caption{($T,\,\mu_B$) values in MeV obtained by fitting the particle ratios using different models.}
\begin{center}
\begin{tabular}{p{0.38in}|p{0.31in}|p{0.31in}|p{0.27in}|p{0.31in}|c|p{0.27in}|p{0.31in}|p{0.31in}|p{0.26in}}
\hline\hline
& \multicolumn{3}{c|}{Ideal gas model} & \multicolumn{3}{|c|}{Rischke model} & \multicolumn{3}{|c}{Present model}\\
\cline{2-4}\cline{5-7}\cline{8-10}
$\sqrt{s_{NN}}$ (GeV) & $T$ & $\mu_B$ & $\delta^2$ & $T$ & $\mu_B$ & $\delta^2$ & $T$ & $\mu_B$ & $\delta^2$\\
\hline
2.70 & 60.0 & 740.0 & 0.85 & 60.0 & 740.0 & 0.75 & 60.0 & 740.0 & 0.87\\
3.32 & 80.0 & 670.0 & 0.89 & 78.0 & 680.0 & 0.34 & 90.0 & 670.0 & 0.69\\
3.84 & 100.0 & 645.0 & 0.50 & 86.0 & 640.0 & 0.90 & 100.0 & 650.0 & 0.60\\
4.30 & 101.0 & 590.0 & 0.70 & 100.0 & 590.0 & 0.98 & 101.0 & 600.0 & 0.53\\
4.85 & 105.0 & 495.0 & 0.30 & 130.0 & 535.0 & 0.84 & 110.0 & 510.0 & 0.43\\ 
8.76 & 140.0 & 380.0 & 0.45 & 145.0 & 406.0 & 0.62 & 140.0 & 380.0 & 0.26\\
12.3 & 148.0 & 300.0 & 0.31 & 150.0 & 298.0 & 0.71 & 148.6 & 300.0 & 0.31\\
17.3 & 160.0 & 255.0 & 0.25 & 160.0 & 240.0 & 0.62 & 160.6 & 250.6 & 0.21\\
130.0 & 172.3 & 35.53 & 0.10 & 165.5 & 38.0 & 0.54 & 172.3 & 28.0 & 0.056\\
200.0 & 172.3 & 23.53 & 0.065 & 165.5 & 25.0 & 0.60 & 172.3 & 20.0 & 0.043\\ 
\hline\hline
\end{tabular}
\end{center}
\end{table}

\begin{figure}[h!] 
\begin{center}
\begin{tabular}{cc}
\begin{minipage}{0.48\textwidth}
\mbox{\includegraphics[scale=0.80]{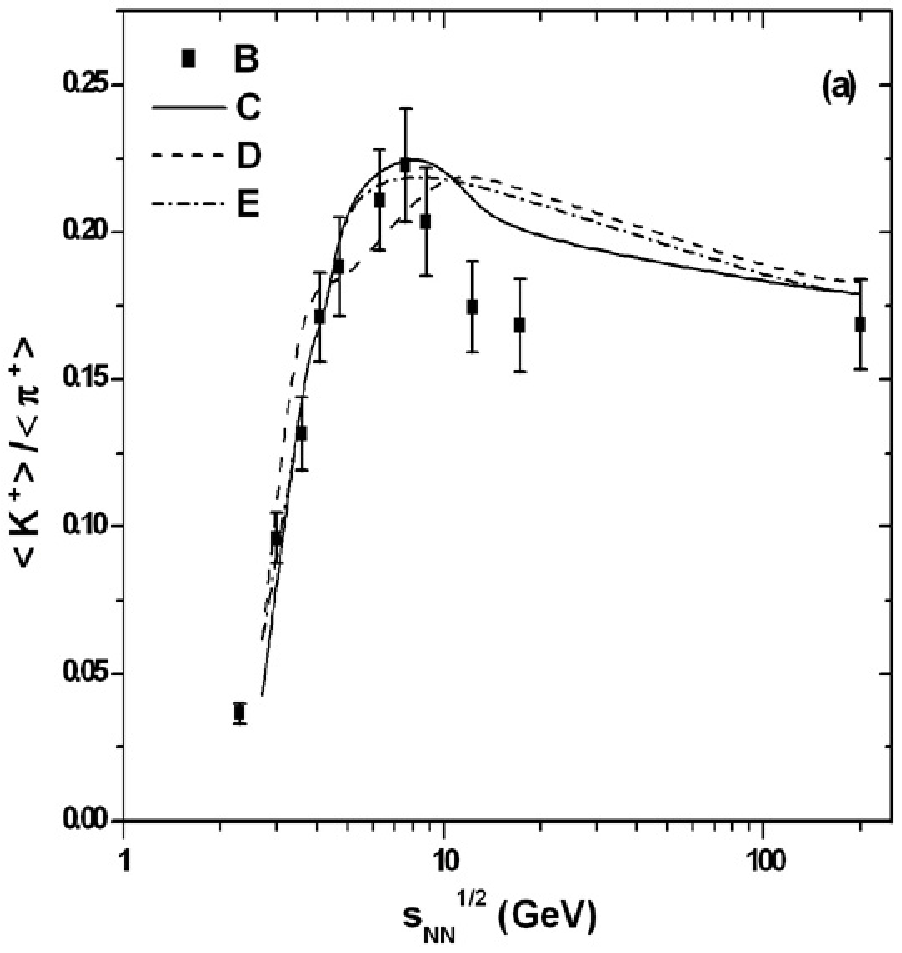}}
\end{minipage}
   &
\begin{minipage}{0.45\textwidth}
\includegraphics[scale=0.80]{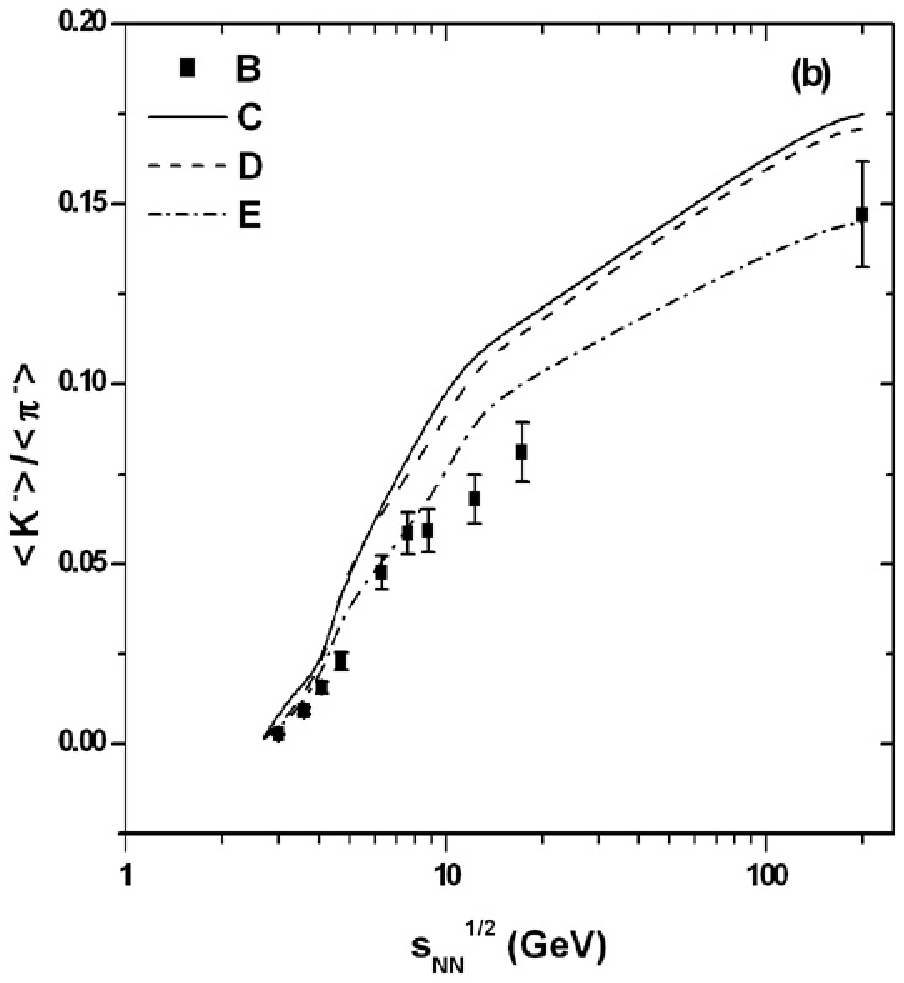}
\end{minipage}
\end{tabular}
\begin{tabular}{cc}
\begin{minipage}{0.50\textwidth}
\includegraphics[scale=0.80]{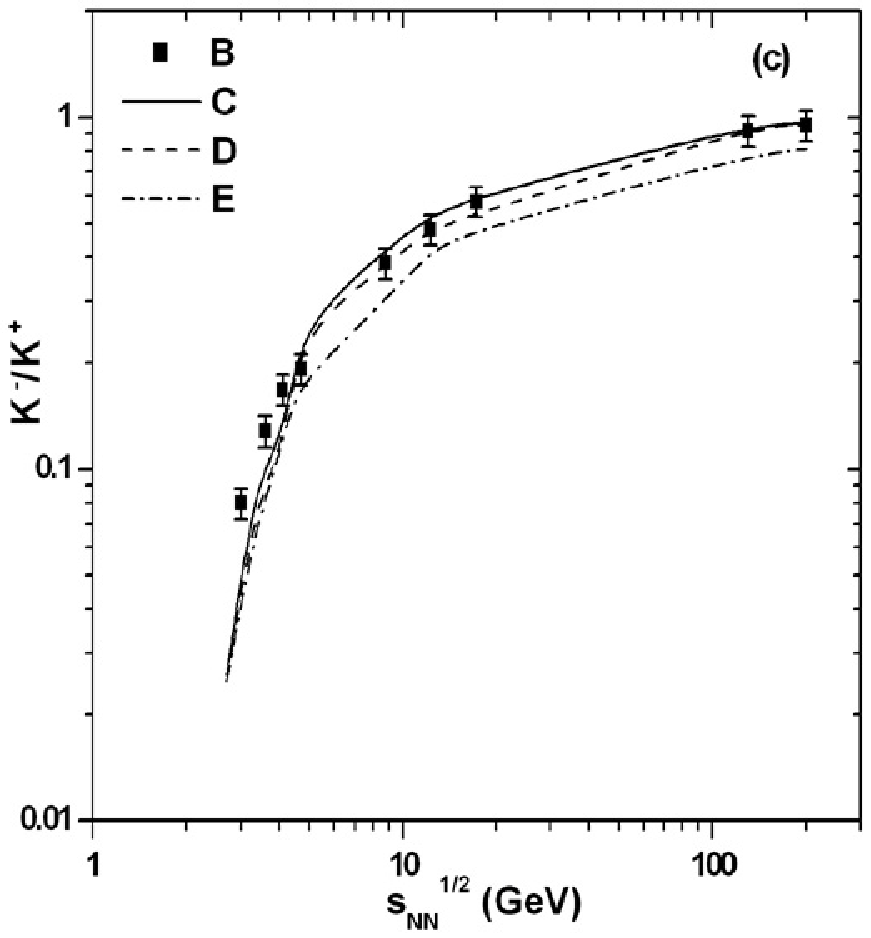}\hskip 0.4cm
\end{minipage} &
\begin{minipage}{0.40\textwidth}
\caption{(a,b,c). Variations of the strange meson to non-strange meson ratios $\langle K^{+}\rangle/\langle\pi^{+}\rangle,\,\langle K^{-}\rangle/\langle\pi^{-}\rangle$ and anti-kaon to kaon ratio $K^{-}/K^{+}$ with respect to center-of-mass energy $\sqrt{s_{NN}}$. Solid curve (C), dashed curve (D) and dash dotted curve (E) show the predictions of the present, ideal HG and Rischke model, respectively. B represents experimental points~\cite{christ}.}
\end{minipage}
\end{tabular}
\end{center}
\end{figure}  

\begin{figure}[h!]
\mbox{\includegraphics[scale=0.77]{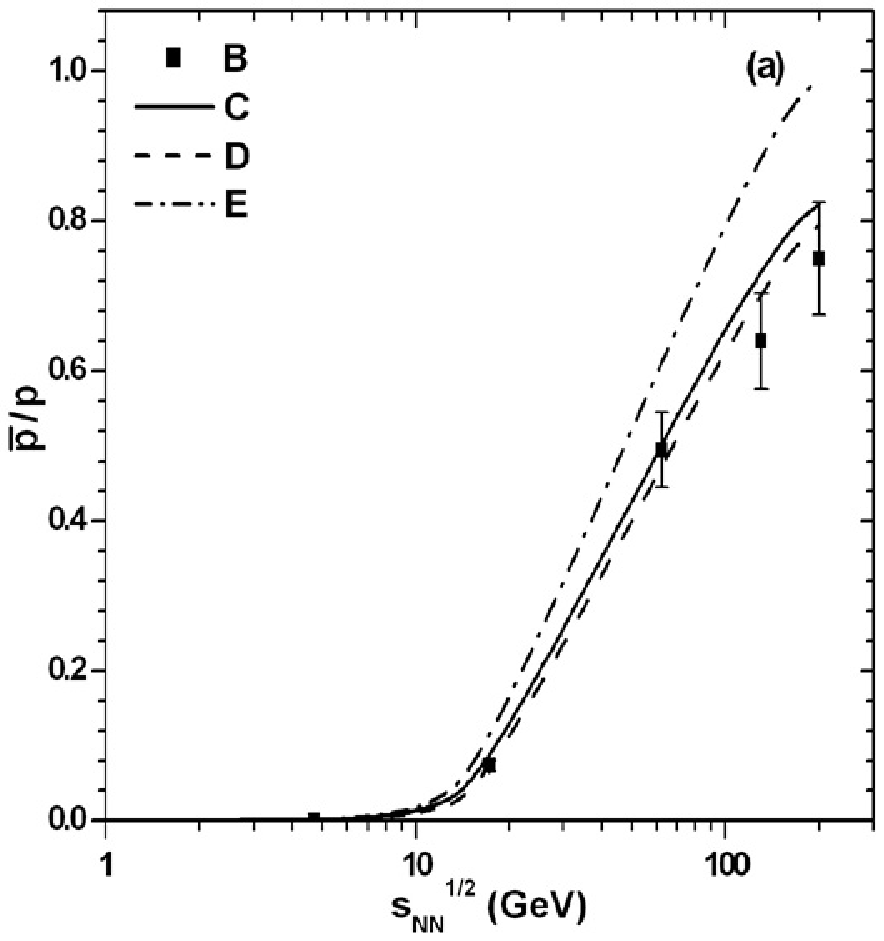}
\hskip -0.50in
\includegraphics[scale=0.77]{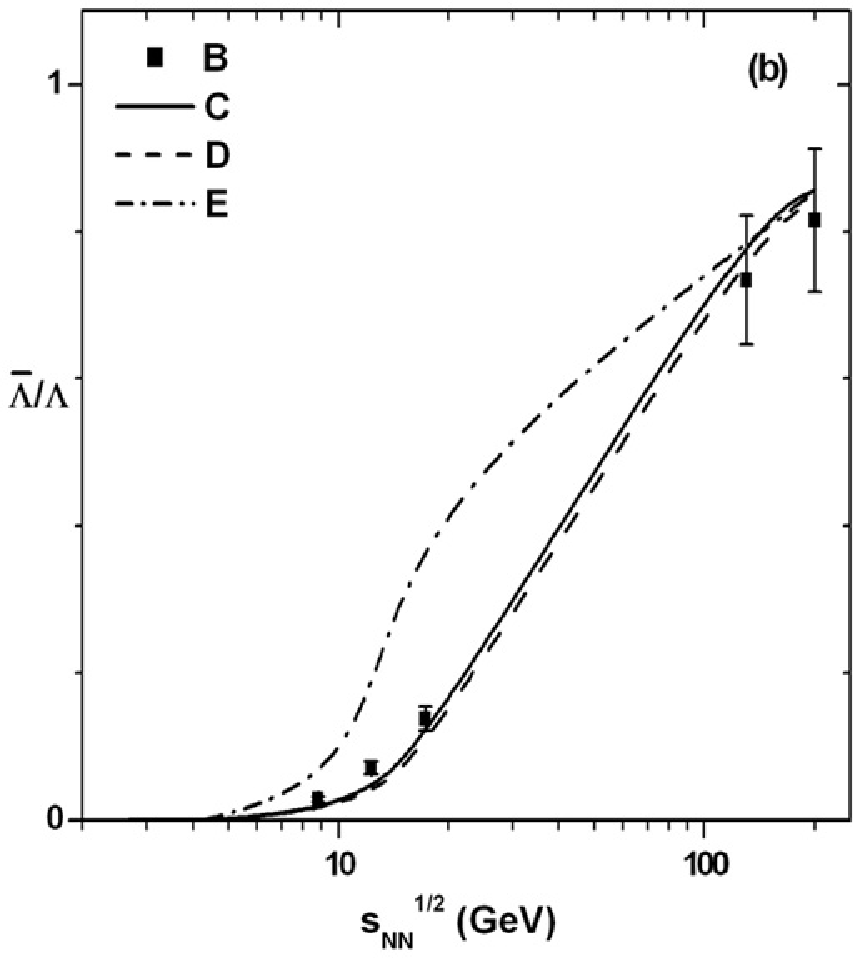}}
\mbox{\includegraphics[scale=0.77]{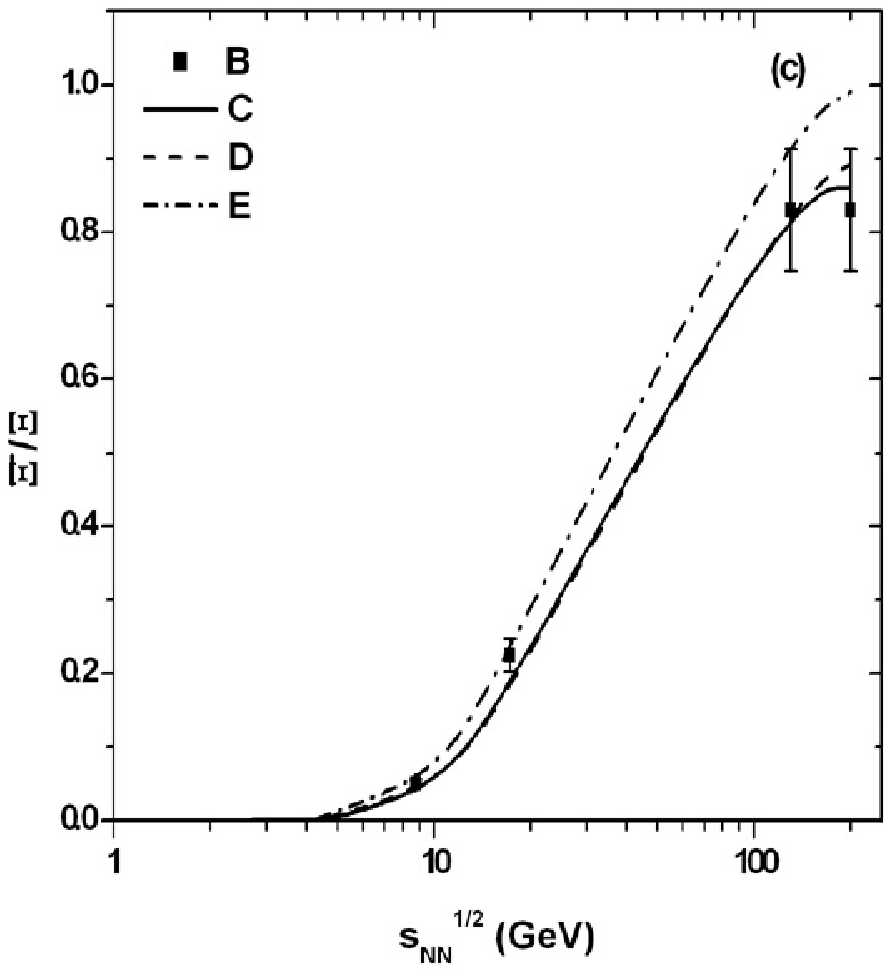}
\hskip -0.50in
\includegraphics[scale=0.77]{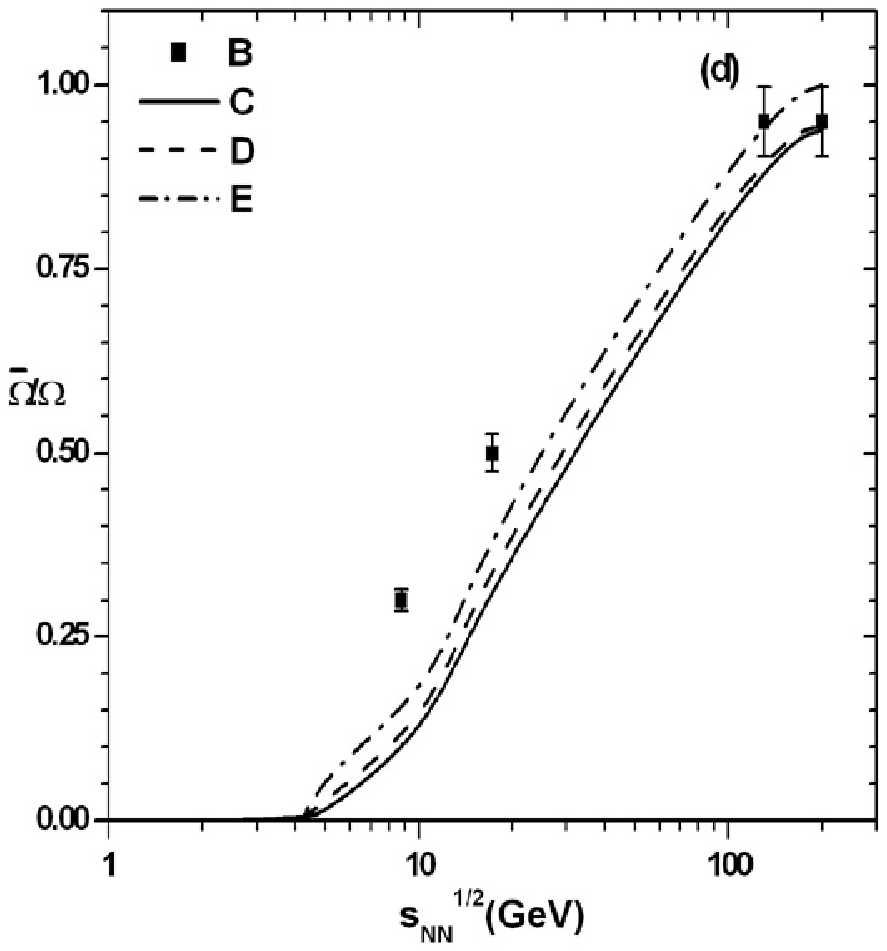}}
\caption{(a,b,c,d). Variations of the anti-baryon to baryon ratios $\bar p/p,\, \bar \Lambda/\Lambda,\,\bar \Xi/\Xi$ and $\bar \Omega/\Omega$ with respect to center-of-mass energy. B represents experimental data whereas curve C, D, and E show the results from present model, ideal HG model and Rischke model, respectively.}
\end{figure} 
                                     
\begin{figure}[h!]
\begin{center}
\mbox{\includegraphics[scale=0.77]{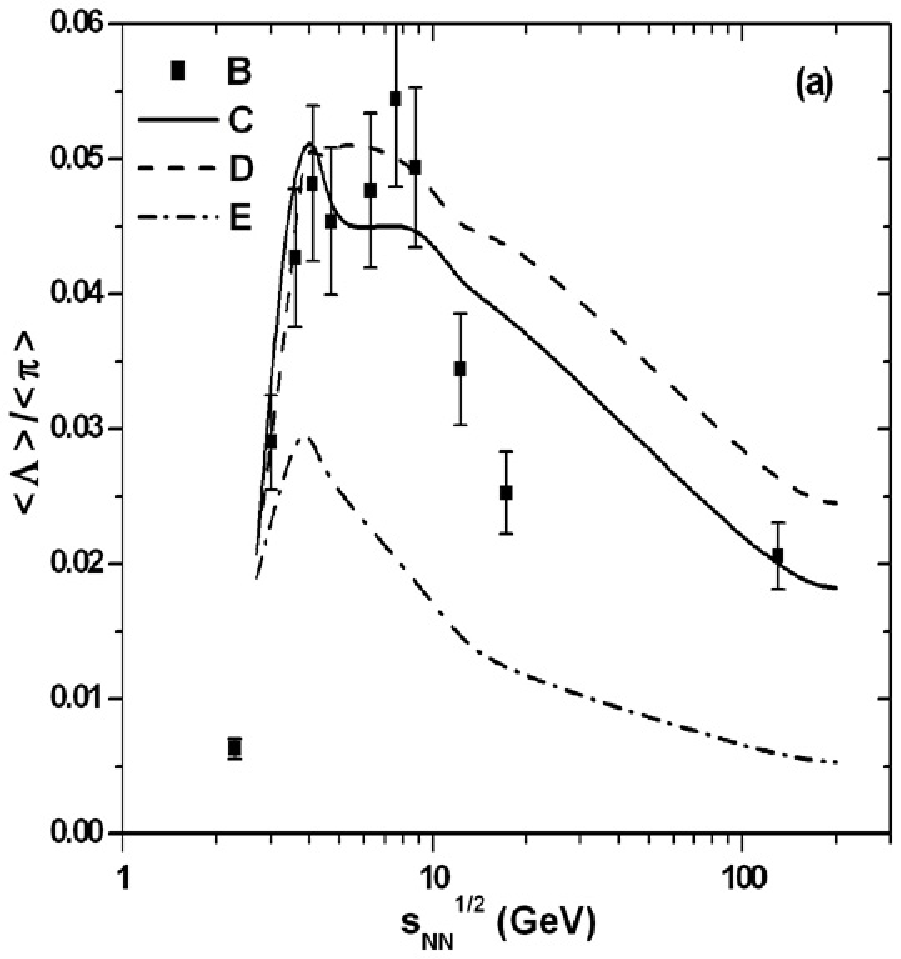}
\hskip -0.50in
\includegraphics[scale=0.77]{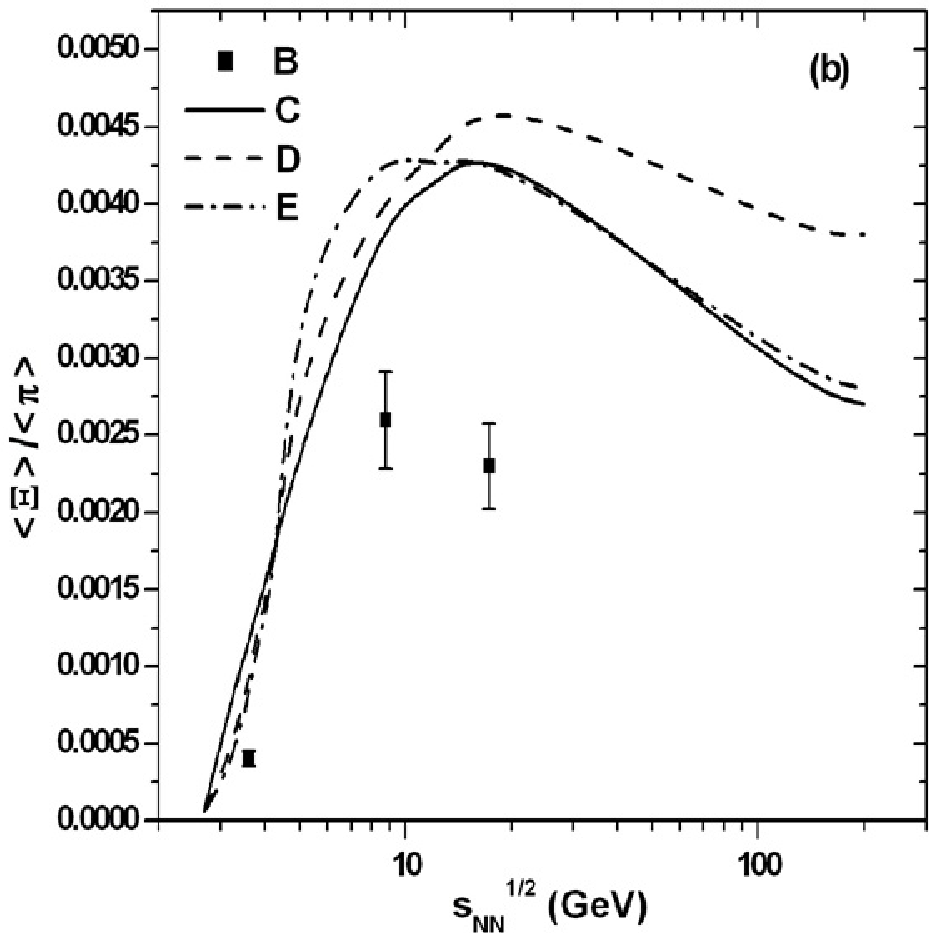}}
\mbox{\includegraphics[scale=0.77]{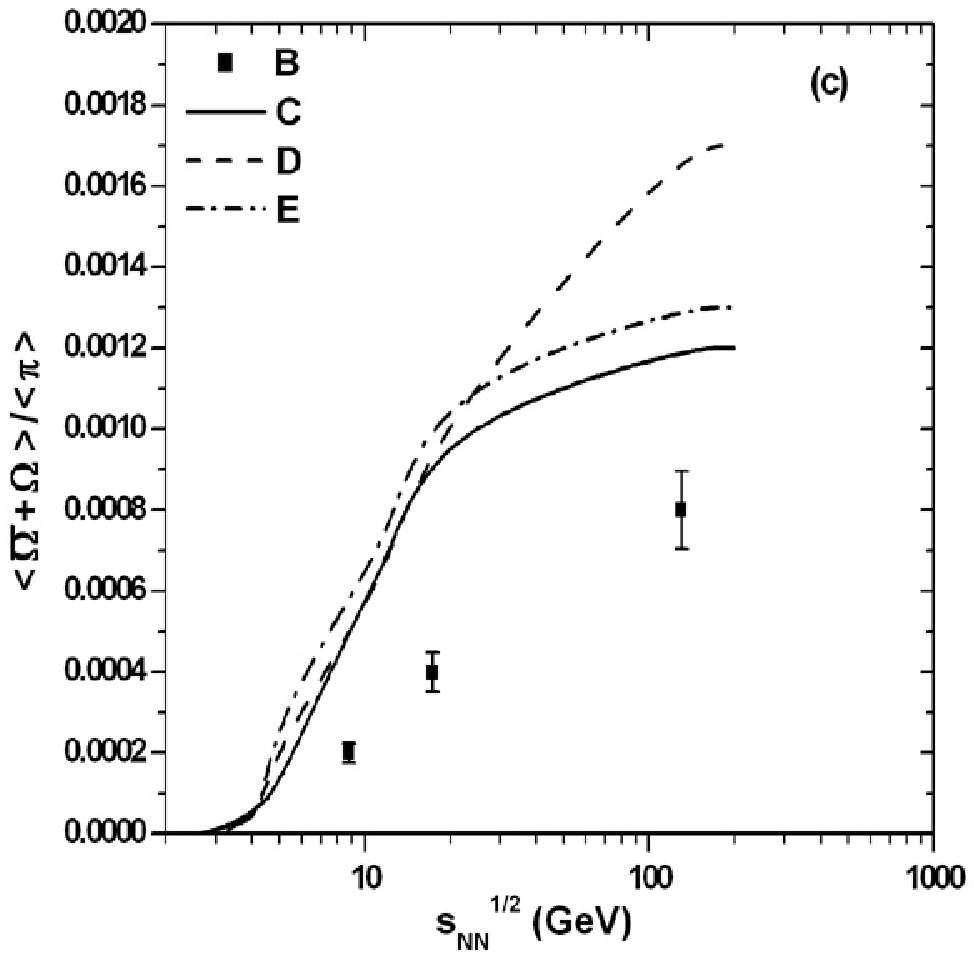}
\hskip -0.50in
\includegraphics[scale=0.77]{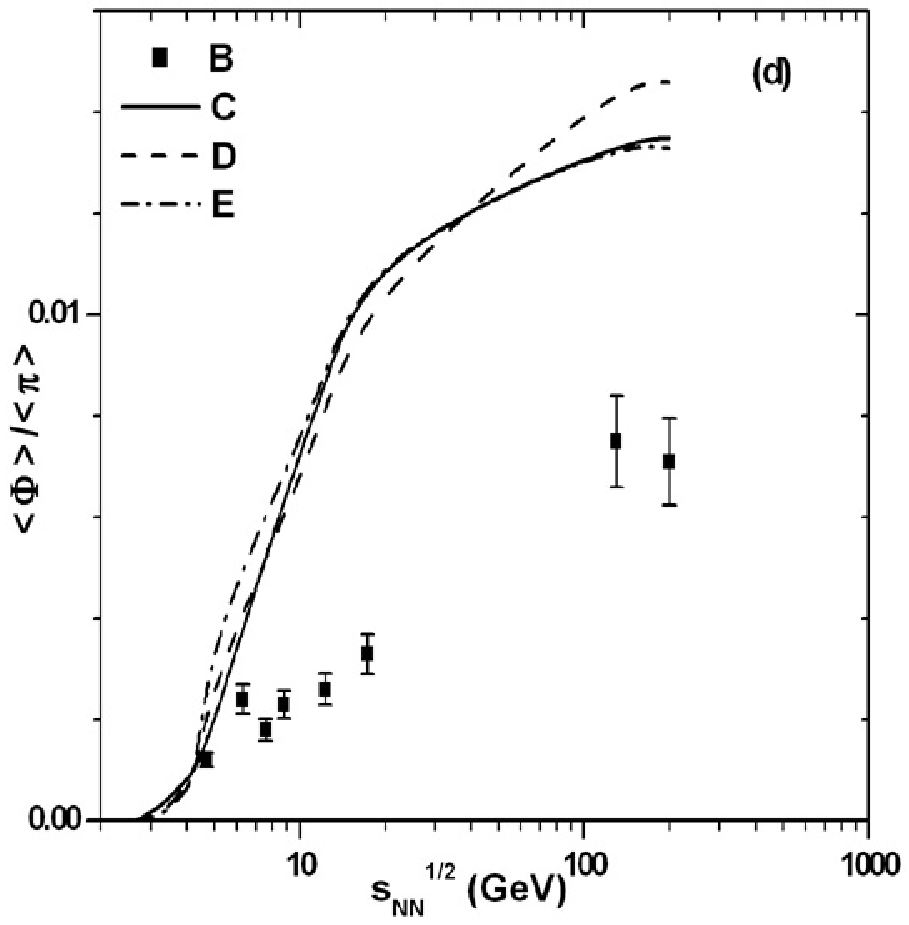}}
\caption{(a,b,c,d). Variations of the strange baryon to non-strange meson ratios $\langle\Lambda\rangle/\langle\pi\rangle,\;\langle\Xi\rangle/\langle\pi\rangle\; {\rm and}\;\langle\Omega+\bar{\Omega}\rangle/\langle\pi\rangle,\;\langle\Phi\rangle/\langle\pi\rangle$ with the center-of-mass energy $\sqrt{S_{NN}}$. In this figure curves B represents experimental data~\cite{christ1} and curves C, D and E depict the predictions of present model, ideal HG model and Rischke model, respectively.}
\end{center}
\end{figure}  
                                                                        
\begin{figure}[h!]
\begin{center}
\includegraphics[height=5.00in,width=5.00in]{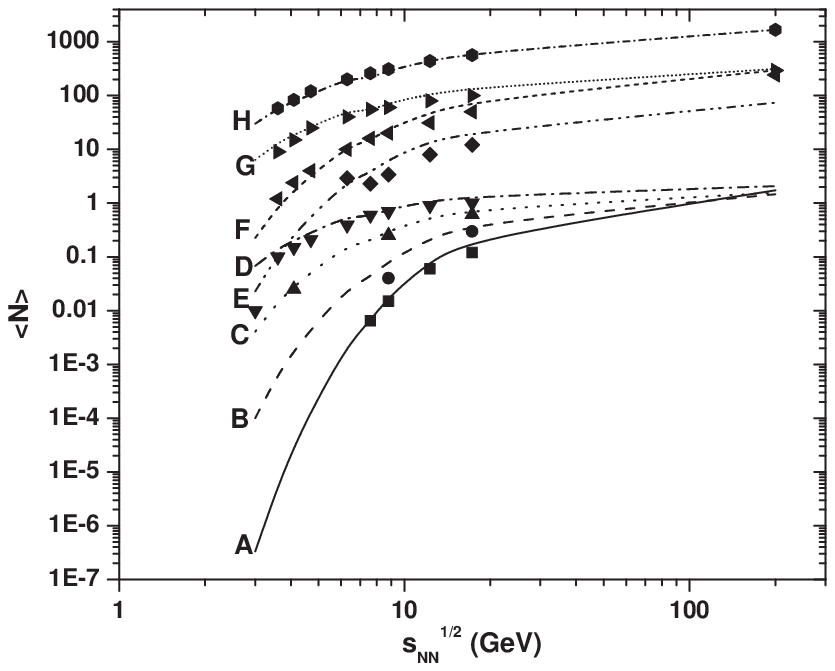}
\caption{Variations of total multiplicities of $\bar{\Lambda},\;(\Omega+\bar\Omega),\;\Xi^{-},\;\Lambda,\;\Phi$, $K^{-},\,K^{+}\; {\rm and}\;\pi^{+}$ with respect to center-of-mass energy predicted by our present model. Experimental data measured in central Au$+$Au/Pb$+$Pb collisions~\cite{klay}-\cite{chung} have also been shown for comparison. In this figure A represents multiplicity of $\bar\Lambda$ (scaled by a factor 0.02), B that of $\Omega+\bar\Omega$ (scaled by 0.2) C that of $\Xi^{-}$ (scaled by 0.1) and D that of $\Lambda$ (scaled by 0.02). Similarly E, F, G and H represent the multiplicities of $\Phi,\,K^{-},\,K^{+},\,\pi^{+}$ mesons, respectively.}
\end{center}
\end{figure}                                   

\section{Results and Discussions}
In order to find out energy dependence of particle ratios and multiplicities we first find energy dependence of chemical freeze-out temperature and baryon chemical potential by fitting the different particle ratios from SIS to RHIC energies using ideal HG model, Rischke model as well as present model. In Table 1, we have given the list of particle multiplicity ratios~\cite{and3} which had been used to calculate $T,\mu_B$ values at freeze-out for different energies. Table 2 summarizes the values of $T$ and $\mu_B$ at different center-of-mass energies in different models. We find that $T$ and $\mu_B$ obtained in our model can be parameterized in terms of center-of-mass energy by using following equations~\cite{oana3,alex3,cley32,jcley99} :
\begin{eqnarray}
\mu_B=\frac{a}{1+b\,\sqrt{s_{NN}}},\\\nonumber
T=c-d\,e^{-f\,\sqrt{s_{NN}}}.
\end{eqnarray}
Here the values of the parameters as arising from the best fit are $a=1.254\pm0.086$ GeV, $b=0.261\pm 0.034$ GeV$^{-1}$ and $c=168.49\pm 4.4$ MeV, $d=171.63\pm 21.75$ MeV $f=0.21\pm 0.039$ GeV.
We compare our values of the above parameters with the values of the parameters $a=1.308\pm0.028$ GeV, $b=0.273\pm0.008$ GeV$^{-1}$ and $c=172.3\pm 2.8$ MeV, $d=149.5\pm 5.7$ MeV $f=0.20\pm0.03$ GeV obtained by Cleymans et al., in ideal HG prescription~\cite{jcley99,cley32}.
We find that the difference in the values of the parameters as obtained by us and by the earlier authors does not appear very significant.  
Recently it has become abundantly clear that the heavy-ion collision energy plays an important role in determining the properties of the final state hadrons. We find that the extracted temperature generally increases rapidly whereas baryon chemical potential monotonically decreases with the collision energy. In general, extracted freeze-out values of these parameters in our model lie close to the ideal gas values. The parameter values in Rischke model are found to differ in comparison to our values.  
   
   Particle ratios and multiplicities are then calculated using these $T$ and $\mu_B$. In order to calculate the multiplicity of hadrons, we first determine the total freeze-out volume by dividing the experimentally measured multiplicities of $K^{+}$ meson at different energies with its number density as calculated in the present model at different center-of-mass energies. We assume that the fireball after expansion, goes through the stage of chemical equilibrium among its various constituents and the freeze-out volume of the fireball for all types of hadrons at the time of the homogeneous emissions of hadrons remains same for all particles.  This freeze-out volume thus extracted, has been used to calculate multiplicities of all other hadrons from corresponding number densities at different center-of-mass energies.
    
   In Fig. 1(a,b,c), we show the center-of-mass energy dependence of the  ratio of strange meson to non-strange meson such as $\langle K^{+}\rangle/\langle\pi^{+}\rangle,\;\langle K^{-}\rangle/\langle\pi^{-}\rangle$ and $K^{-}/K^{+}$ in our present model, ideal hadron gas model and Rischke model. Solid, dashed and dash dotted curves represent predictions of our model, ideal hadron gas model and Rischke model, respectively. The $\langle K^{+}\rangle/\langle\pi^{+}\rangle$ ratio shows a peak at around 8.0 GeV of center-of-mass energy. While the ratio $\langle K^{-}\rangle/\langle\pi^{-}\rangle$ shows a monotonic increase with center-of-mass energy $\sqrt{s_{NN}}$ and achieves a saturation at or around RHIC energy. Essentially the saturation arises as the freeze-out temperatures that become almost constant at or around RHIC energies. 
  The peak in the $\langle K^{+}\rangle/\langle\pi^{+}\rangle$ ratio is demonstrated in all the thermal models~\cite{braum,jcleyma} and this successful explanation of the experimental data is one important feature of the thermal models. However, the peak in these models is slightly broader than that observed by experiments. Our thermal model shows a peak at the same energy as measured in the experiment and we find a good agreement between our thermal model prediction and the experimental results~\cite{christ}. Other types of models like transport model etc. are not able to explain the experimental feature of $\langle K^{+}\rangle/\langle\pi^{+}\rangle$ ratio. However, all types of thermal models provide a large ratio for $K^{-}/\pi^{-}$ than observed in experiments. Still the main feature of the data showing a steady increase  is also reproduced by all thermal models. The ratio $K^{-}/K^{+}$ shows  much dependence on the center-of-mass energy. Our thermal model prediction for the ratio $K^{-}/K^{+}$ shows a close agreement with the predictions of other models and also matches with the experimental results. The feature of the experimental curve is also in a good agreement with our thermal model calculation. 
        
    In Fig. 2(a,b,c,d), we plot the energy dependence of the anti-baryon to baryon ratios like $\bar p/p,\;\bar\Lambda/\Lambda,\;\bar\Xi/\Xi$ and $\bar\Omega/\Omega$ as given by our present model, ideal hadron gas model and Rischke model. Surprisingly we find that our present model and the ideal hadron gas model reproduce the qualitative and quantitative features of the experimental data for all of these ratios. This occurs because the freeze-out values of $T$ and $\mu_B$ at different energies do not differ much in these two models. However, the predictions from the calculations in Rischke model do not show such agreement with the experimental data. As we have demonstrated in the previous chapter that ideal HG description does not provide a proper EOS for hot and dense hadron gas. The successful explanation of the data shows that a proper and realistic EOS for HG is given by our model which gives a thermodynamically consistent description of the hot and dense HG.   
 
    Fig. 3(a,b,c,d) show the variation of the particle ratios $\langle\Lambda\rangle/\langle\pi\rangle,\;\langle\Xi\rangle/\langle\pi\rangle$ and $\langle\Omega+\bar{\Omega}\rangle/\langle\pi\rangle,\;\langle\Phi\rangle/\langle\pi\rangle$ with the center-of-mass energy $\sqrt{s_{NN}}$ in all the above three models. We find that results in our present calculation particularly of $\langle\Lambda\rangle/\langle\pi\rangle$ show good agreement with the experimental data~\cite{christ1} since our curve reproduces the main feature of the data. Ideal hadron gas model also gives a satisfactory explanation but results lie far below the experimental points. This again supports our earlier claim that a suitable description for HG is given by the EOS in our model. Main problem comes when we compare predictions from all these models for multi-strange particle productions e.g.,$\langle\Xi\rangle/\langle\pi\rangle$ and $\langle\Omega+\bar{\Omega}\rangle/\langle\pi\rangle,\;\langle\Phi\rangle/\langle\pi\rangle$ as shown by Fig. 3(b,c,d). We find that in all such cases, thermal model predictions do not match with the experimental data and in fact theoretical curves lie far above the curves given by the experimental data. Here $\Xi$ is $ssu$ (or $ssd$), $\Omega$ is $sss$ and $\Phi$ is given by $s\bar s$ quark combinations. These results not only signify the failure of all types of thermal models but also emphasize the need for going beyond such approaches for the cases of multi-strange hadrons. The strangeness enhancement signals QGP formation but in these cases, thermal models yield much larger strangeness than observed by the experiments. However, it has been pointed out by several authors~\cite{vgre3,rchw3,rjf3,dmol3,dmol31,lwch3} that such anomalous results are compatible with quark-coalescence models invoking a QGP formation.     
    
    In Fig. 4, we show the center-of-mass energy dependence of multiplicities of hadrons $\bar{\Lambda},\;(\Omega+\bar\Omega),\;\Xi^{-},\;\Lambda,\;\Phi$, $K^{-},\,K^{+}\; {\rm and}\;\pi^{+}$ as predicted by calculations in our present model. Curves A, B, C and D show the multiplicities of $\bar\Lambda$, $\Omega+\bar\Omega$, $\Xi^{-}$ and $\Lambda$ baryons scaled by factors of $0.02$, $0.2$, $0.1$ and $0.02$, respectively. Curves E, F, G and H depict the multiplicities of $\Phi$, $K^{-}$, $K^{+}$ and $\pi^{+}$ mesons, respectively. We have also shown here experimental results measured in central Au$+$Au/Pb$+$Pb collisions~\cite{klay}-\cite{chung} for comparison. Here we have extracted freeze-out volume of the fireball from the calculated number density of $K^{+}$ and comparing it with total multiplicity of $K^{+}$ experimental data. We use the same volume for all other particles. We observe an excellent agreement between the theoretical predictions by our present thermal model and the experimental data for the total multiplicities of $\pi^{+},\,K^{+},\,K^{-},\,\Lambda,\,\bar\Lambda$ etc. However, thermal model calculation slightly differs for $\Omega+\bar\Omega,\,\Xi$ and $\Phi$ as compared to experimentally measured values. Thermal values of the multiplicities for all these particles are again larger than the experimental values. This analysis again suggests a new and different mechanisms for the production of these particles. One way out of this difficulty is to assume that these particles achieve chemical equilibrium earlier in the fireball when the corresponding volume is much smaller. However, this is a vexing problem appearing in the use of thermal models and we have to find appropriate answers to these problems.   
       
\section{Summary and Conclusions}
We have formulated an excluded-volume hadron gas model and used it in order to analyze the variations of multiplicity of various particles with respect to the center-of-mass energy $\sqrt{s_{NN}}$ from SIS to RHIC energies. We have also used the present model to explain the variations of some particle ratios again with center-of-mass energy and compared our results with the experimental data. A good agreement between our present model results and experimental data supports the claim that thermal model gives a satisfactory description of the data.

   In the past we have witnessed the success of ideal gas model in explaining the data, but we know that a correct description of hot, dense hadron gas can be given by a model where hard-core repulsive interactions are incorporated in a thermodynamically consistent way. Moreover, at GSI-SIS and BNL-AGS energies, the freeze-out parameters involve a much larger values of baryon chemical potentials and the predictions of all the excluded-volume models are quite different from those obtained in the ideal gas models. We have already shown the utility of the present model in explaining various properties of hot, dense hadron gas~\cite{cp,cp1}. The analysis presented here lends further support to our claim that the excluded-volume model obtained by us properly explain the multiplicities and particle ratios of various particles after chemical freeze-out. In conclusion, our model provides a proper and realistic EOS for a hot, dense hadron gas and it can successfully be used at extreme values of the temperatures and/or densities.   
     
\section*{Acknowledgments}
   One of the authors (M. Mishra) would like to acknowledge the financial support from Council of Scientific and Industrial Research (CSIR), New Delhi. C. P. Singh acknowledges financial support in the form of a Department of Science and Technology (DST) project from Government of India.  

\end{document}